\documentclass[onecolumn,showpacs,preprintnumbers,prb,aps]{revtex4}

\usepackage{graphicx}
\usepackage{dcolumn}
\usepackage{amsfonts,amsmath,amssymb,bm}
\usepackage{natbib}
\usepackage{hyperref}
\setcitestyle{super}

\begin{document}

\title{\bf {Effect of Ag doping on structural, optical, and photocatalytic properties of ZnO nanoparticles}}
\date{\today} 
\author{S.~M.~Hosseini}
\author{I.~Abdolhosseini~Sarsari*}
\author{P.~Kameli}
\author{H.~Salamati}
\affiliation{a) Department of Physics, Isfahan University of Technology, Isfahan, 84156-83111, Iran}

\newcommand{\etal}{{\em et al}}

\begin{abstract}
Silver-doped ZnO nanoparticles were successfully fabricated 
at  $400\,^{\circ}\mathrm{C}$ via a simple and rapid method based on 
short time solid state milling and calcination of precursor powders.
The effect of Ag dilute doping on the 
structural, optical, and photocatalytic properties of 
ZnO nanoparticles was investigated by X-ray diffraction 
(XRD), UV-vis spectrophotometer and photoluminescence (PL) spectroscopy.~X-ray analysis
revealed that Ag doped ZnO solidified in hexagonal wurtzite structure.~
The intensity of deep level emission was reduced with increasing silver 
doping in PL measurement.~The X-ray photoelectron spectroscopy (XPS) 
measurement predicted that Ag was mainly in the metallic state and ZnO was in the wurtzite structure.~
This metallic state accompanied by unique zinc oxide properties decolorized the methyl violet, efficiently.~
The first-principles calculation represented Ag deep level in ZnO with an n-type behavior, while in ZnO structure with 
grain boundary p-type nature via shallow states is dominant same as powder samples as studied in this present work.
It was suggested that these Ag-doped ZnO nanoparticles may have good 
applications in optoelectronics, spintronics and wastewater treatment.
\end{abstract}

\maketitle

\section{INTRODUCTION}
Multifunctional materials come on stage accompanied with magnetic properties in diluted
magnetic semiconductors (DMS), in which non-magnetic ions replace some of the magnetic and non-magnetic cations.
The origin of ferromagnetic properties in diluted semiconductors has been studied, both theoretically and 
experimentally~\cite{Das2008, Sarsari2013}, however, still there is no broad consensus on this issue .

Among DMS materials, ZnO-based DMS has
remarkable features:~i) Direct wide band gap $3.37~eV$ at room
temperature,~ii) Access to the large bulk and high-quality single crystals 
of ZnO and iii) Large exciton binding energy of $60~meV$.~The latter is useful 
for having lasers with high efficiency at room temperature~\cite{tsu2004, ozg2005}.
In addition, ZnO is useful in solar cells and sensors as transparent conductive and
piezoelectric~\cite{ah2009}.

Pure ZnO nanostructures show weak optical features that are resulted from point defects 
such as oxygen vacancy or interstitial Zn; therefore, they can not be used directly in
the industry~\cite{ah2009}.~As a result, doping ZnO with a convenient element is a
method for engineering optical and magnetic properties.~Furthermore, in order to make 
optoelectronic devices, n-type and p-type states are needed.~It 
is relatively difficult to achieve p-type doping and this limitation is 
considered an obstacle in using ZnO in these instruments~\cite{tek2011}.~
Although the elements V, I, and IB groups have been studied for p-type doping~\cite{sahu2014},
silver has been reported as the best candidate because of its high solubility, larger ionic size,
and minimum orbital energy~\cite{yan2006}.~
Theoretically, Volnianska et al.~\cite{Vol2009} reported that, among the elements of group IB, copper, and especially
Au, ionization energy was higher than Ag and caused silver to be the best case for
achieving p-doping in ZnO.~
Recently, Thomas et al.~\cite{thomas2012} combined experimental 
approaches and theoretical calculations and revealed that the majority of Ag was doped (substitution) into 
the ZnO structure. Furthermore, Ag 4d and O 2p states were overlapped to form an impurity 
band, which shifted the Fermi level toward the valence band maximum and induced p-type 
properties in ZnO. Electrical properties of the Ag-doped ZnO nano-wires showed that their 
p-type properties depended not only on Ag content, but also on other conditions 
such as electrochemical growth and post-annealing conditions. 

Silver ions have two characteristic features and can be
used in the places of substitution and interstitial; hence, they can act as an acceptor in ZnO 
~\cite{lupan2010}. However, previous studies on Ag-doped ZnO by Yan et al. 
~\cite{yan2006} suggested that substitutional sites were more energetically
favourable than interstitial sites.
Most studies have examined the effects of silver doping on the photocatalytic activity and
antibacterial properties of ZnO~\cite{Karun2011, amo2012, chen2008,Kumar2015}.~It is also 
possible to form a dilute magnetic semiconductor by silver doping in ZnO wurtzite structure
and ferromagnetic behavior of the compound has been investigated experimentally~\cite{he2011}
and theoretically~\cite{li2012}, which can be used in spintronic applications by 
both degrees of freedom of electron spin and charge.~The ferromagnetic properties are 
due to the formation of impurity band which interacts with the d orbital of nonmagnetic
element~\cite{coey2005}.~However, the origin of ferromagnetic properties is still under investigation.

In the studies on the nanostructures of Ag-doped ZnO, different methods such as
sol-gel~\cite{rad2013}, sonochemical~\cite{karunak2010}, and coprecipitation~\cite{chau2010} have been used, 
all of which require harsh conditions of sample preparation, high temperature, and high costs.
However, in this paper the structural, optical, and photocatalytic properties in $Zn_{1-x}Ag_{x}O$ 
($x=0, 0.02$, and $0.07$) nanoparticles 
synthesized by a simple and low-cost method, called thermal treatment of ball-milled precursors were investigated.
The aim of this work was to produce p-type doping of the nanoparticles 
to be used in practical applications.~Optical measurements showed that the synthesized nanoparticles were of p-type.

\section{EXPERIMENTAL}

\subsection{Substances and their provision}
Nanoparticle samples of $Zn_{1-x}Ag_{x}O$ were prepared by a simple and
low-cost method.~In this method, zinc acetate dehydrate $[(CH_{3}COO)_{2}Zn*2H_{2}O]$ with
$95\%$ purity and citric acid $[C_{6}H_{8}O_{7}]$ were initially combined at the molar ratio of $1:1$.
Then, various concentrations (x=$0$, $0.02$ and $0.07$  called A, B, and C, respectively) of silver nitrate $[AgNO_{3}]$ were used for 
preparing the doped samples.~The powders were mixed and ground for 90 min at room
temperature and the milled powders were ultimately calcined in air at $400\,^{\circ}\mathrm{C}$ for 2 h~\cite{zandi2011}.

\subsection{Photocatalytic studies}
The photocatalytic performances of the samples were evaluated by the decomposition of  
methyl violet in the solution under UV light. In order to prepare the samples for the photocatalytic 
analysis, first,  0.05 gr of~MV with $ C_{24}H_{28}N_{3}Cl$ chemical formula was weighed. 
The 50ppm solution of methyl violet dye and 0.1 gr of the catalyst,~ZnO, and Ag-doped ZnO,
 in a quartz tube were prepared for the photo reactor system. First, the solution was placed in the dark mood 
 for 30 min and then, under UV~irradiation (wavelength=254 nm, power =15 W) at several stages.~
 Afterwards, absorption spectra of the samples were taken. The solution was analyzed by 
 Lambda 25 UV/VIS spectrophotometer ( Perkin Elmer).

The degradation rate was calculated using the following formula: 
\begin{equation}
Degradation~percent = \frac{C_{0}-C(t)}{C_{0}}\times 100(\%)
\end{equation}
where $C_{0}$ is the primal concentration of methyl violet solution and C(t) is 
concentration at different UV irradiation times.~

\subsection{Characterization of samples}

Structural analysis was carried out using X-ray diffractometry (XRD) (XPERT model, $Cu K_{\alpha}$ radiation, $\lambda=1.5405\AA$
at room temperature). The morphologies of the prepared nanoparticles were recorded
 by a Hitachi S4160 field emission scanning
electron microscope (FE-SEM). X-ray photoelectron spectroscopy (XPS) analysis was carried out for assessing
purity of materials and examining the composition and characteristics of defects by an ESCA/AES system. 
The system was equipped with a concentric hemispherical analyzer (CHA, Specs model EA10 plus). 
An $Al K\alpha$ line at 1486.6 eV was utilized for exciting the X-ray photoelectrons. This analysis was done
using high-resolution scan of the O(1s), Zn(2p), Zn(3d) and Ag(3d) spectral regions. In the XPS spectra, 
all the binding energies were calibrated by taking the carbon peak (284.8 $eV$) as the reference.~
Formation of ZnO wurtzite phase and available molecular bonds was investigated by the measurement of 
FTIR absorption spectrum of $Zn_{1-x}Ag_{x}O$ samples using a~BRUKEN TENSOR 27 spectrophotometer. 
The FTIR analysis was carried out at room-temperature in the range 400-4000 $cm^{-1}$ using KBr as the reference.
To investigate the optical properties of these nanoparticles, the absorbance spectra of the samples were measured 
on a Perkin Elmer Lambda 25 UV/VIS spectrophotometer using ethanol as the reference. The UV-vis absorption 
spectrum was recorded in the range of 300-800 nm. Room temperature photoluminescence spectra of the samples 
were acquired from Perkin-Elmer LS55 spectrometer using xenon flash lamp laser as the excitation source within the 
range of 340 to 500 nm and excitation wavelength of 210.5 nm.

\section{RESULTS AND DISCUSSION}

\subsection{Structural properties}
Fig.~\ref{XRD} shows a broad and enlarged view of X-ray diffraction
patterns of the $Zn_{1-x}Ag_{x}O$ nanoparticles.~
All the intensity data were appropriately normalized to 1.
The XRD pattern of pure ZnO
displayed wurtzite crystal structure (JCPDS NO.~01-079-2205), without any
other crystalline phases.~There were three additional peaks at $2\theta$ of 
38.14, 44.48, and 64.47$^{o}$ in ZnO doped with Ag values 0.02 and 0.07, which could be 
attributed to metallic Ag fcc phase, and indicate the formation of Ag as the 
second phase clusters.~Also, a consistent increase in the intensity of silver
peaks can be noted by the increase in the concentration of Ag
from 2 to 7 $mol\%$.~It should be noted that the formation of
metallic silver and the crystalline phase of ZnO begins at about $400\,^{\circ}\mathrm{C}$~\cite{Geo2008}. 
~By increasing 
Ag concentration, peak position is shifted toward lower values, as shown in Table~\ref{XRD}.~
This shift suggests the partial substitution
of $Ag^{+}$ ions at the ZnO lattice and increase in the lattice parameters a and
c, as expected~\cite{Kar2011}; this issue is presumably related to the  ionic size difference between the
$Ag^{+}$ (0.126 nm) and $Zn^{2+}$ (0.074 nm) ions.~But, in this work,
rietveld analysis showed a slight decrease in the a and c lattice parameters and 
cell volume, which can be explained by the formation of silver 
nano-collection~\cite{Vo2009} and the part substitution of $Ag^{+}$ ions in the system,
respectively.~The substitution did not cause any increase in the lattice parameters
via very small percentage of silver doping.~   

To study the effect of silver doping on the crystal size,
the mean crystal size was estimated from the Debye-Scherrer
equation as follows:

\begin{equation}
D=\frac{k\lambda}{\beta\cos \theta}
\end{equation}

where, $\beta=FWHM$, $\theta$ is Bragg angle, $\lambda=1.54\AA$ (X-ray wavelength) and $k=0.9$.~
The crystalline size of A, B and C samples are estimated as 23.12, 24.44 and 25.46 nm, respectively.

\begin{figure}
\begin{center}
\includegraphics*[scale=0.3]{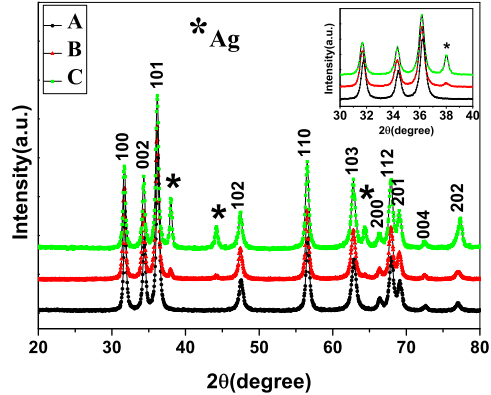}
\caption{\label{XRD}
(color online) XRD patterns of A, B and C samples.~The inset of
it pointed the enlarged view of XRD patterns by $2\theta=30-40$.}
\end{center}
\end{figure}
\begin{table}
\begin{center}
\caption{\label{XRD}
Geometric parameters of un-doped and Ag-doped ZnO nanoparticles.
}
\begin{tabular}{|c|c|c|c|}
\hline
Compound &  2Theta  & (hkl) & Crystal size "D" (nm) \\
\hline
A    & 31.78 & (100) & 23.12\\
\hline
B   & 31.67 & (100) & 24.44\\
\hline
C   & 31.65 & (100) & 25.46\\
\hline
\end{tabular}
\end{center}
\end{table}

\begin{figure}
\begin{center}
\includegraphics*[scale=1.5]{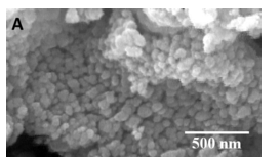}\\
\includegraphics*[scale=1.5]{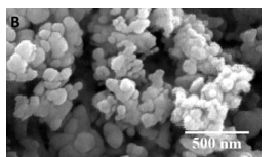}\\
\includegraphics*[scale=1.5]{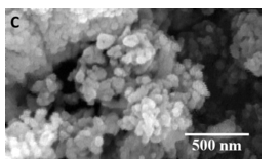}\\
\caption{\label{FE-SEM}
The FE-SEM morphologies of (a) undoped, (b) $2\%$ and (c) $7\%$ Ag doped ZnO nanoparticles.
}
\end{center}
\end{figure}

Information such as grain size and shape can be obtained by 
FE-SEM analysis.~Surface morphology of the undoped and silver-doped ZnO nanoparticles is shown in Fig.~\ref{FE-SEM}.~
The particle sizes were clearly of the order of nanometers and shape of
particles was quasi-spherical.~Formation of the  quasi-spherical nanoparticles in the size range of 20-150 nm is shown in  
Fig.\ref{FE-SEM}.~Using silver doping, significant changes were not 
observed in the size of nanoparticles.

\subsection{FT-IR studies}
\begin{figure}
\begin{center}
\includegraphics*[scale=0.3]{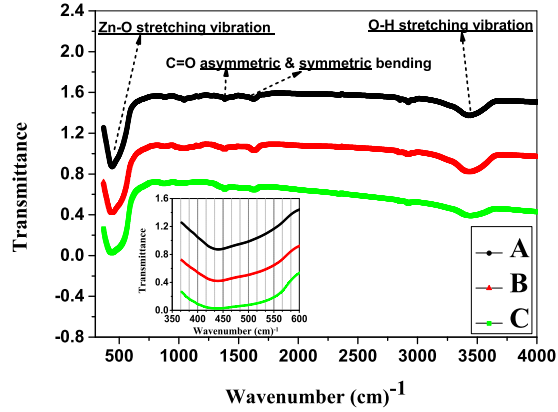}
\caption{\label{FTIR}
FTIR spectra for undoped and silver doped ZnO.~the inset shows the
enlarged view of first absorption band location.
}
\end{center}
\end{figure}

FT-IR spectra for different concentrations of silver doping in the range of 300-4000 $cm^{-1}$ were  
recorded and the results can be seen in Fig.~\ref{FTIR}. Enlarged position of the first absorption 
band is shown in the inset.~Sharp peak was observed in 440 $cm^{-1}$ , which could be  attributed to the Zn-O 
stretching vibration mode~\cite{Babu2014}.~A wide peak was in the range 
of 3020-3650 $cm^{-1}$ that was related to the presence of hydroxyl ions (OH) in the ZnO:Ag system.~
The symmetric and asymmetric bending modes of C=O bonds were in 1580 and 1410 $cm^{-1}$, while the 
 peak located at 2860 and 2950 $cm^{-1}$ was related to symmetric and asymmetric C-H stretching bonds, 
 respectively~\cite{zandi2011}. There were some bands originated from the presence of water moisture and carbon 
 dioxide in the air in the process of making pellet. The absorption band at 1020 $cm^{-1}$ could be attributed to
 bending vibrational modes~\cite{lanje2013}. According to Fig.~\ref{FTIR}, a slight shift with increasing 
 silver (Ag) is visible. The shift in the position of the band toward lower frequencies can be associated with 
 changes in bond length due to the partial substitution of $Ag^{+}$ ion at the ZnO lattice~\cite{Mur2014}.

\begin{table*}
\caption{\label{xps}
Binding Energy$(eV)$ and Area size of $Zn(2p_{1/2})$, $Zn(2p_{3/2})$, $O(1s) (I)$, $O(1s)
(II)$, $Ag(3d_{3/2})$ and $Ag(3d_{5/2})$ in XPS analysis of morphology of ZnO and A and C samples.~
}
\begin{tabular}{ccccccc}
                        &\multicolumn{6}{c}{Binding Energy$(eV)$@ Area}\\ \cline{2-7}
       Sample    &$Zn(2p_{1/2})$&$Zn(2p_{3/2})$&$O(1s) (I)$&$O(1s) (II)$&$Ag(3d_{3/2})$&$Ag(3d_{5/2})$\\ 
 \hline
A  &1044.81@1.179    &1021.67@1.835     &530.01@0.863     &531.75@0.906   &-                            &-\\
C &1044.56@1.296    &1021.55@1.801     &529.64@0.604     &531.38@1.235   &372.66@0.707 &366.52@0.616 \\
\end{tabular}
\end{table*}

\subsection{XPS results}
\begin{figure}
\begin{center}
\includegraphics*[scale=0.3]{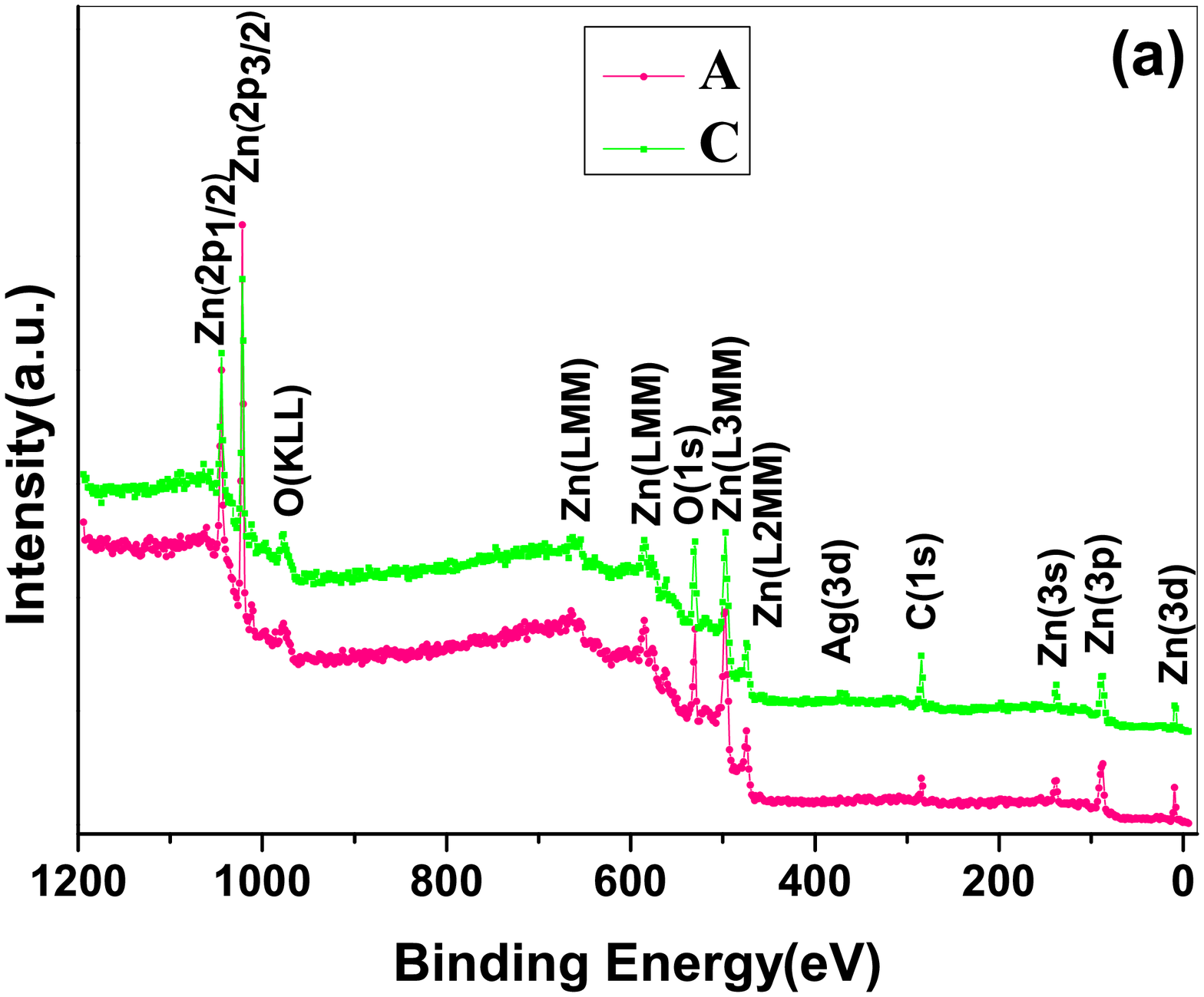}\\
\includegraphics*[scale=0.3]{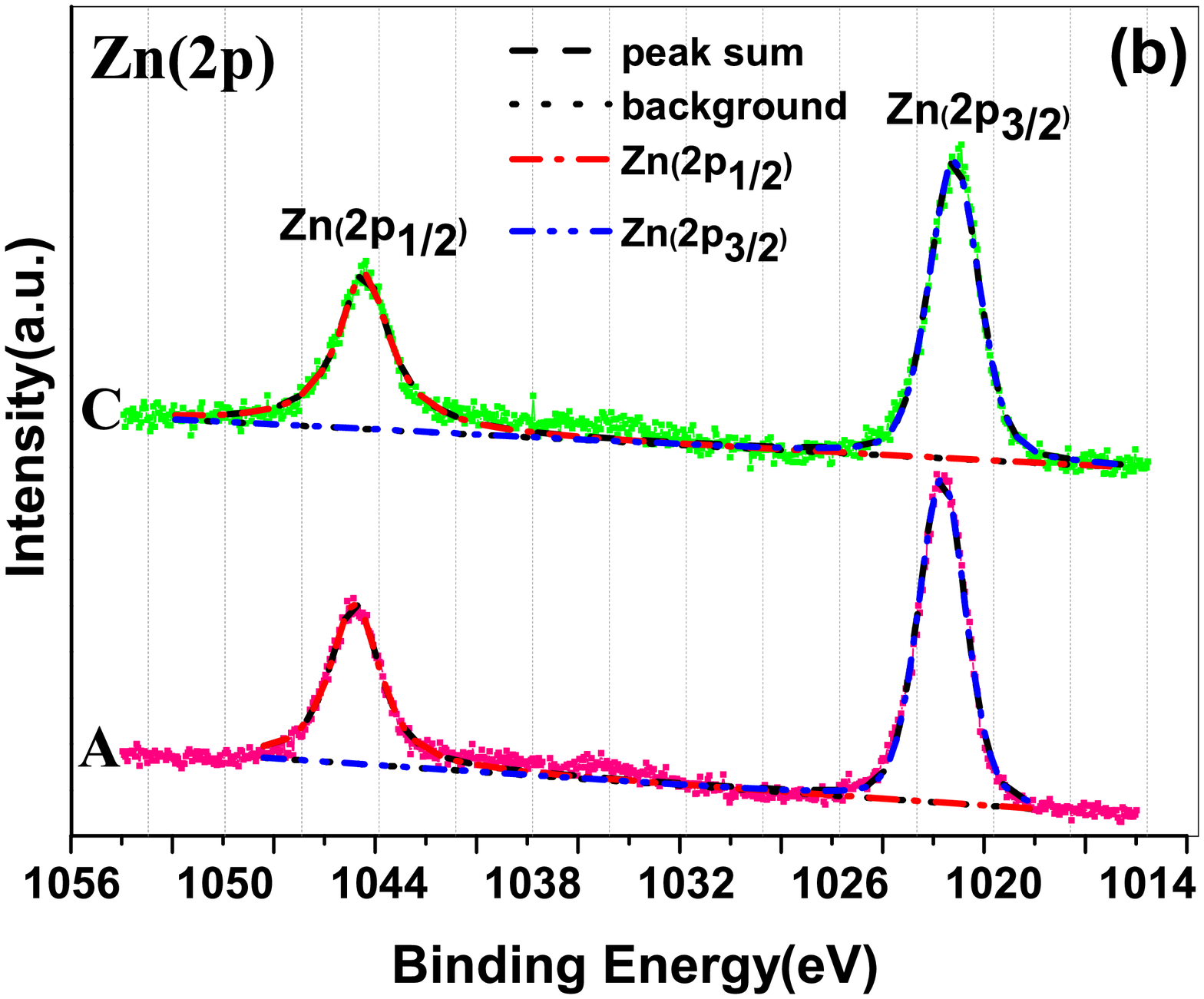}\\
\includegraphics*[scale=0.3]{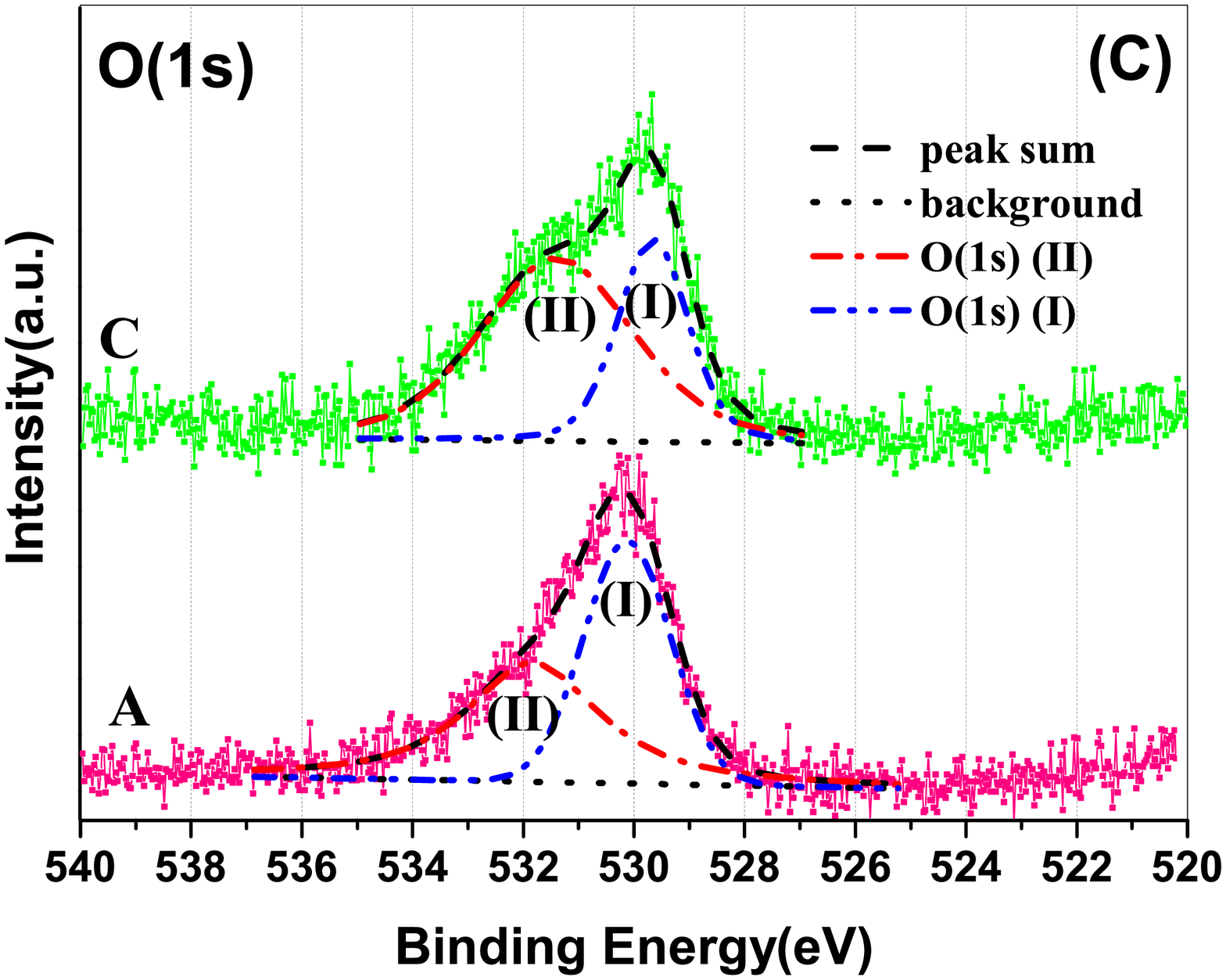}\\
\includegraphics*[scale=0.3]{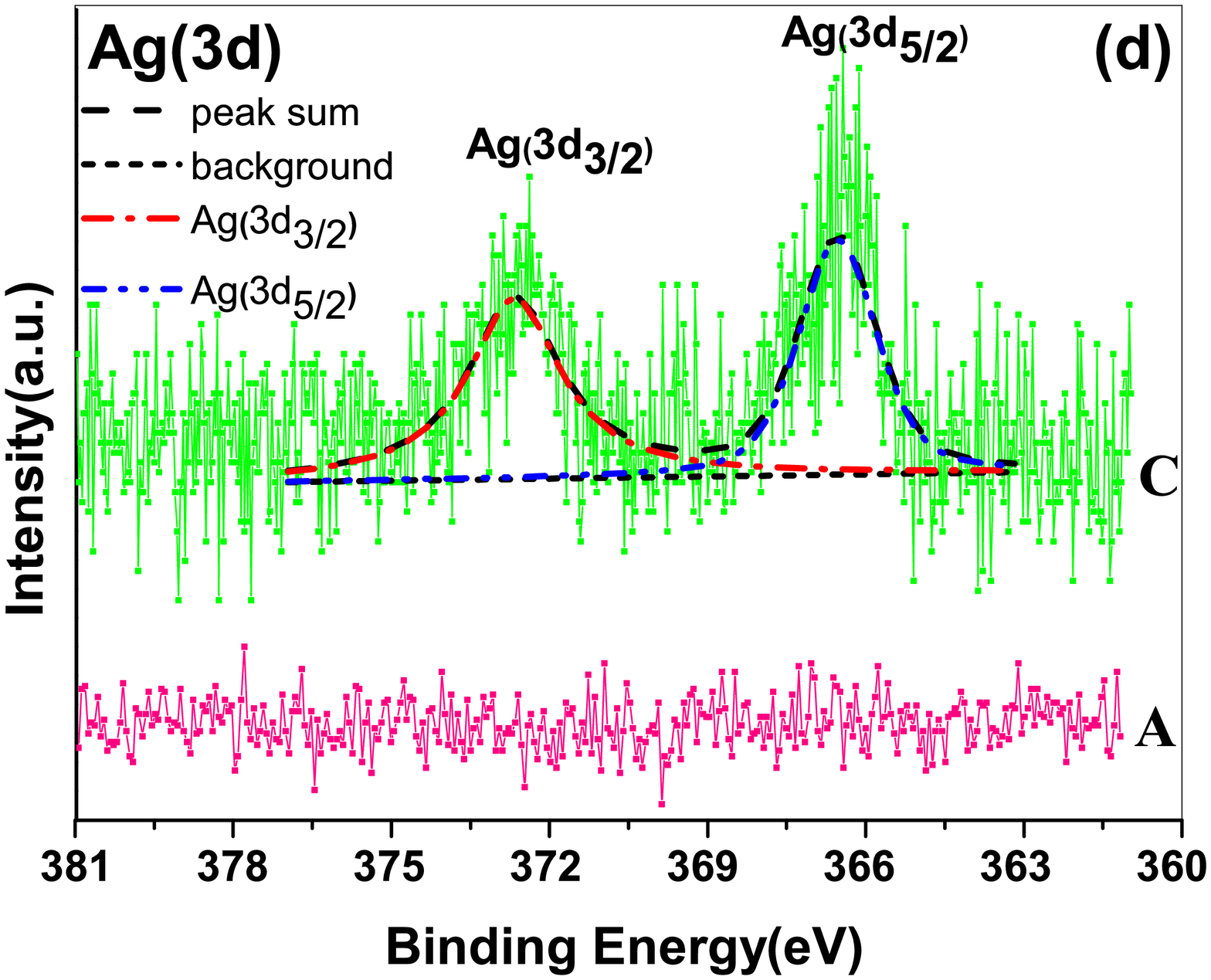}\\
\caption{\label{XPS}
(Color online) XPS spectra of A and C samples: 
(a) The survey scan spectrum; (b) Zn(2p) spectrum; (c) O(1s) spectrum and (d) Ag(3d) spectrum.
}
\end{center}
\end{figure}

To determine the elements and chemical-bonding state of the compounds in the $Zn_{1-x}Ag_{x}O$ nanoparticles, 
the XPS analysis of the samples was studied.~It should be noted that the XPS curves were fitted using XPSPEAK 
4.1 software.~The survey and core-shell level XPS spectra of Zn-2p, O-1s, and Ag-3d as well as valance band of ZnO and
 $7\%$ silver-doped ZnO  
nanoparticles are shown in Fig.~\ref{XPS}. In the survey scan spectrum, Zn, O, C, and Ag peaks were characterized, as can be seen in Fig.~\ref{XPS} (a). The carbon present in the samples was presumably due to acetate vestige~\cite{Habibi2010} and/or carbon adsorption process present in ambient conditions~\cite{AlG2013, wahab2007}. 

Symmetrical peaks are visible in Fig.~\ref{XPS}(b) located at 1044.81 and 1021.67 $eV$ for the sample of pure 
zinc oxide ascribed to $Zn(2p_{1/2})$ and $Zn(2p_{3/2})$, respectively~\cite{Jin2011}. The splitting of Zn-2p 
states was about 23 $eV$ which was induced from the powerful spin-orbit coupling. These numbers were different from
 the binding energy quantity of stoichiometric ZnO (1045.1 $eV$ for $Zn(2p_{1/2})$ and 1022.1 $eV$ for $Zn(2p_{3/2})$), 
 which can be attributed to the change of charge transfer from $Zn^{2+}$ to $O^{2-}$ due to the existence of vacancies~\cite{Sah2012}.~
 According to Table~\ref{xps} and Fig.~\ref{XPS}(b), the location of peaks was shifted toward the lower binding
 energy by silver doping. These peaks could confirm the wurtzite structure and $Zn^{2+}$ mode of Zn atom on the samples
 ~\cite{Meng2013}. Therefore, due to the charge transfer in the binding energy of XPS spectrum, it can be deduced that the 
 Zn-2p peak shift indicates that oxygen deficiency is the dominant factor versus Zn deficiency in the doped samples
 in contrast to the pure one~\cite{Sah2012}.

Fig.~\ref{XPS}(c) shows the internal levels of oxygen in both of the nanoparticles.
It is clearly seen that the O-1s curve was asymmetric; hence, they were fitted with two 
Gaussian peaks (as I and II) symmetrically.~There were two different types of oxygen
groups in the samples.~The  peak (I) was located in the lower binding energy in comparison
with the peak (II), which was dedicated to $O^{2-}$ ions of $Zn-O$ bonding at crystal
lattice~\cite{Zheng2011}.~The peak (II) was related to the hydroxyl group adsorption
due to structural defects~\cite{AlG2013}; these groups could play a major role in the
photocatalytic activity and increase the photocatalytic activity~\cite{Chong2010}
because of preventing from the recombination of electron-hole pairs.~Silver
doping in ZnO led to a shift in the O-1s spectrum toward lower
binding energy.~The atomic ratio of hydroxyl group (II) to total oxygen was 
calculated for pure ZnO, which was equal to $51\%$ and increased by 
$16\%$ using Ag doping.~So, the photocatalytic activity should
be increased by silver addition process.~Also, decreasing the binding energy of
the Zn-2p and O-1s levels with silver doping could be due to the
reduction of oxygen vacancy.~Decrease in the ratio of  $Zn/O$ in the present XPS results
for the doped samples compared to the pure sample confirmed the reduced the 
density of oxygen defects~\cite{zheng2007}.

The chemical state of Ag element was characterized by examining Ag-3d levels in Fig.~\ref{XPS}(d). 
The $Ag (3d_{3/2})$ and $Ag (3d_{5/2})$ peaks were split to about the $\sim$6 $eV$ that showed the 
metallic condition of silver in the present synthesised samples~\cite{Ao2008, wu2010}. Lupan et al.~\cite{lupan2010} 
and Khosravi et al.~\cite{khosravi2014} have revealed the presence of two different components, which could be 
attributed to either metallic Ag or $Ag_{2}O$ and to a Ag-Zn-O ternary compound. Interestingly, the binding 
energy of Ag-3d for the doped sample compared to the binding energy of silver bulk sample (368.3 $eV$ 
 for $Ag (3d_{5/2})$ and 374.3 $eV$ for $Ag (3d_{3/2})$~\cite{zheng2008}) was significantly shifted towards 
 lower binding energy. Thongsuriwong et al.~\cite{thong2012} observed a similar behavior in Ag 3d spectra for 
 Ag-doped ZnO thin films. This shift showed the electron movement from metallic Ag levels to ZnO nanoparticles 
 and the formation of silver with unit valance. The interaction between silver and ZnO nanoparticles 
 was also proposed, which could regulate the position of Fermi level of silver and zinc oxide nanoparticles and 
 result in the formation of a new Fermi level for metallic silver. Free electrons above the new Fermi 
 level might be tunnelled to the conduction band, because the conduction band of ZnO was empty~\cite{zheng2007}. 
 As mentioned before, in the present XPS patterns, Ag-3d binding energy was shifted to lower binding energy which can be explained by 
 two processes: the first process is that the binding energy of the unit valance silver is much lower than the zero 
 valance silver~\cite{Habibi2010} and the second one is due to the formation of an oxide layer on the surface of silver
 particles~\cite{Sah2012}, which can be identified by the XRD analysis. Due to the high electronegativity of Ag than Zn, electron 
 transfer was assumed to occur from Zn to the Ag particles, suggesting that the chemical bond between ZnO and Ag is one 
 reason for silver effect in controlling defects at ZnO crystal lattice.

\subsection{Optical studies}
\begin{figure}
\begin{center}
\includegraphics*[scale=0.25]{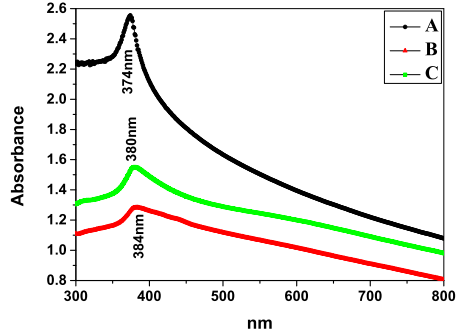}
\caption{\label{UV1}
UV-visible absorption spectra of pure and doped ZnO nanoparticles.
}
\end{center}
\end{figure}

Optical absorption spectra of the $Zn_{1-x}Ag_{x}O$ samples were measured by UV-vis
spectrophotometer in the range 300-800 nm.~UV-vis absorption spectrum
primarily represents the available change in the energy gap~\cite{pal2010}.
Fig.~\ref{UV1} shows the UV-vis spectra of the ZnO:Ag nanoparticles.~Pure ZnO nanoparticles indicated
the strong UV absorption edge at 374 nm that could be related to the wurtzite crystal 
structure of ZnO~\cite{wahab2007}.~Silver adding to the ZnO nanoparticles
moved the absorption edge toward the major wavelength (red shift).~Changes in the
absorption peaks, due to doping, showed changes in the band structure.~
The shift of the absorption edge represented the change in the particle 
energy gap~\cite{mittal2014}.~

The optical energy gap of the nanoparticles was obtained from drawing the graph of $(Ah\nu)^{2}$ 
versus energy, according to $A=\frac{[k(h\nu-E_{g})^{\frac{n}{2}}]}{h\nu}$
equation.~In this relation, A is absorption, k is constant, h is Planck's constant, $\nu$ is 
light frequency, n = 1 for direct electronic transition, n = 4 for indirect electron transition 
and $E_{g}$ is energy gap~\cite{jen2012}.

The A sample's optical energy gap was 3.25 $eV$ that was smaller than the
energy gap of the bulk samples of ZnO.~With increasing the amount of silver, the energy
gap decreased from 3.25 $eV$ for the pure sample (x=0) to 3.18 $eV$ for the Ag-doped sample (x=0.07), as shown in 
Table~\ref{UV}.~The increase of the energy gap for ZnO doped with donor is commonly 
observed, while reducing the energy gap of ZnO doped with acceptor has been reported~\cite{sinha2011}.
Decreasing the energy gap can be related to the presence of p-type conductivity in
the silver-doped ZnO nanoparticles.~Silver doping in ZnO provides the impurity band in the 
energy gap, which could be due to the formation of the p-type in this substance.~It should
be mentioned that this reduction in energy gap led to increase efficiency in the use of these materials in
optoelectronic devices~\cite{sinha2011}.

\begin{table}
\begin{center}
\caption{\label{UV}
UV–vis absorption data for the undoped and doped ZnO nanoparticles.
}
\begin{tabular}{|c|c|c|}
\hline
Sample &  $\lambda(nm)$  & Band gap$(eV)$ \\
\hline
A    & 374 & 3.25\\
\hline
B   & 384 & 3.21\\
\hline
C   & 380 & 3.18\\
\hline
\end{tabular}
\end{center}
\end{table}

Further information about the optical properties
of ZnO can be received from PL spectra.~Information about the crystal modality,
structural defects such as oxygen vacancy, Zn interstitials,...~and surface 
properties of particles can be obtained from PL spectra~\cite{Geo2008}.

\begin{figure}
\begin{center}
\includegraphics*[scale=0.27]{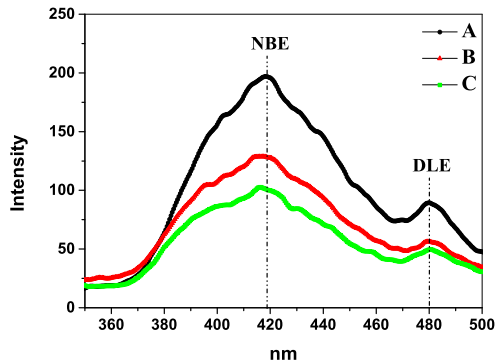}
\caption{\label{PL}
(color online) PL spectra of the pure ZnO and Ag doped ZnO nanoparticles.~
}
\end{center}
\end{figure}

Fig.~\ref{PL} shows the PL spectra of ZnO doped with silver and pure ZnO nanoparticles. 
A radiation band at the center of the UV spectral range around 418 nm for all the samples 
can be observed. The UV radiation is usually attributed to the near band edge (NBE) 
emission due to free exciton recombination~\cite{Jin2011}. No significant change was 
observed in the position of this ultraviolet band with silver doping. The decrease in UV 
intensity was due to the interactions between the excited ZnO nanoparticles and Ag particles 
in the grain boundaries. These kinds of interactions via the Schottky contact, metal-semiconductor 
diode effect, decrease the recombination of electrons and holes generated from UV light 
irradiation~\cite{Du2014, Geo2008} and improved the photocatalytic activities~\cite{liq2006}.

On the other hand, reduction in the intensity of UV radiation was due to the interstitial Ag atoms 
at ZnO lattice and created a large amount of defects, as confirmed by Zeferino et al.~\cite{zeferino2011}. 
Dissimilar results for the annealing effect on the PL properties of Ag-doped ZnO nanowires were 
reported by Yong et al.~\cite{yong2013}. They observed the increase in UV intensity that was 
attributed to the substitution of Zn atom sites by Ag atoms. The present experimental results in 
Fig.~\ref{PL} displayed deep level emissions (DLE) at about 480 nm in the visible region. 
This multicomponent visible emission in ZnO has been frequently ascribed to several intrinsic 
and extrinsic defects. DLE or blue radiation is due to electron recombination in oxygen vacancy 
with a hole in the valence band~\cite{Kar2012}. By increasing Ag concentration, the intensity of DLE 
decreased. The reduction of DLE with Ag doping represented a decline in the defects and improved 
crystallization ZnO~\cite{Karun2011,Jin2011}.

\begin{figure}
\begin{center}
\includegraphics*[scale=0.27]{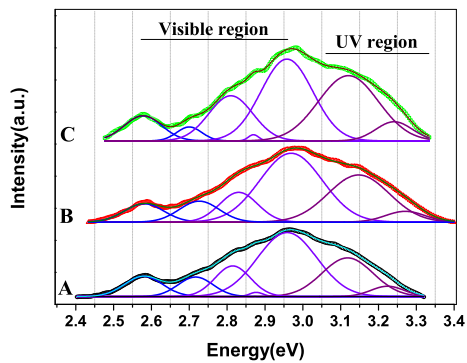}
\caption{\label{fit}
(color online) Gaussian fitting of the A, B and C samples in the visible and UV region.
}
\end{center}
\end{figure}

To better understand the role of Ag doping in the optical properties, Gaussian fitting was performed on PL~spectra.
Multiple PL peaks in the energy region of the absorption edge were observed in many ZnO nanostructures. 
Understanding the origin of the PL peak was required to study the optical properties of ZnO nanostructures and 
sketch ZnO-based opto-electronics tools. PL spectra's Gaussian fitting for pure and silver-doped ZnO nanoparticles 
is shown in Fig.~\ref{fit}. The first band on 2.58 $eV$ as a blue irradiance was reported due to the electron 
transition between Zn interstitials and Zn vacancy states~\cite{Xu2013, zeferino2011} which was in agreement with the 
calculated energy levels of defects in ZnO by Kohan et al.~\cite{kohan2000} and showed the existence of 
Zn vacancies in the samples, while oxygen defects constitute deep donor levels in the band gap~\cite{Jay2013}. 
Band located at 2.74 $eV$ was blue radiation that was attributed to the Zn vacancy defects~\cite{Lin2013}. 
Radiant bands on 2.95, 2.88, and 2.81$eV$ were violet radiation.
 The mechanism of this radiation can be questioned: the emission mechanism is still unknown and researchers 
 have related it to defects of Zn interstitials or Zn vacancy defects~\cite{Xu2013}. The two peaks located at 
 3.24 and 3.12 $eV$ were UV near absorption edge radiation due to free exciton recombination.
The NBE emission in UV region was shifted which represented the variation of the band gap.

\subsection{ Studying photocatalytic activity }
The purpose of this analysis was the degradation of methyl violet.~In the industry,
methyl violet is entered as a pollutant into groundwater and its destruction helps to 
remove pollutants.~When ZnO nanoparticles are used to remove indissoluble chemicals
in diverse applications, surface properties such as surface area, oxygen vacancy, and
hydroxyl ions play an important role~\cite{wang2004}.

\begin{figure}
\begin{center}
\includegraphics*[scale=0.3]{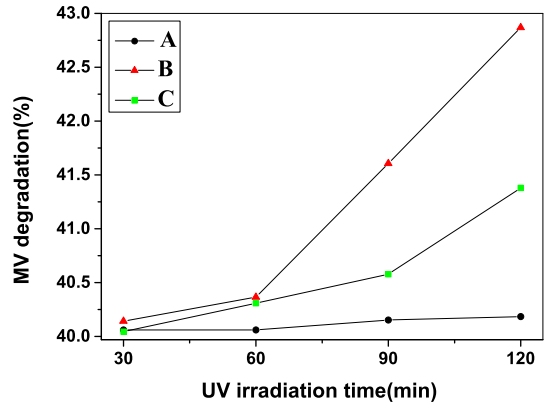}
\caption{\label{photo}
(Color online) MV degradation curves of A, B and C samples under UV irradiation.
}
\end{center}
\end{figure}

The results of photocatalytic measurements of the ZnO:Ag nanoparticles are shown
in Fig.~\ref{photo}.~Pure ZnO nanoparticles showed a significant photocatalytic activity.
It can be seen that the photocatalytic activity of the ZnO nanoparticles was
improved by silver doping with optimum 2 $\%$, which may be related to the oxygen
 vacancy defect concentration in different samples.
In the C sample compared with the B one,
the Zn vacancies were more
dominant than the oxygen vacancies due to Ag presence on grain boundaries and consequently 
lack of Zn atoms in the bulk region.~So, increasing Zn vacancies accompanied 
by oxygen vacancies could be responsible for the degradation process.
According to these studies, it can be concluded that the photocatalytic activity of the ZnO
nanoparticles increases by Ag doping and thereby incurs the increased surface 
area and increased oxygen defects.~The reaction between conduction-band 
electrons and oxygen in the solution could generate the reactive oxygen species which is responsible
for the color decolorisation~\cite{amo2012} and its value increases with silver doping.~As a result, 
the increase of silver improves the photocatalytic activity of the ZnO  nanoparticles .

\begin{figure}
\begin{center}
\includegraphics*[scale=1.8]{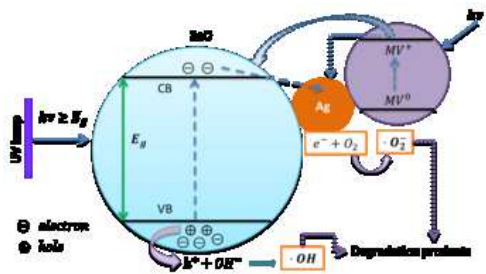}
\caption{\label{photocatalytic}
Schematic mechanism of photocatalysis process on ZnO:Ag nanoparticles under UV irradiation.
}
\end{center}
\end{figure}

Fig.~\ref{photocatalytic} schematically explains the mechanism of photocatalytic process on ZnO:Ag nanoparticles.~
In the dark, by dispersing ZnO:Ag nanoparticles in the MV solution, the surface electrons of the Ag nanoparticles were transferred
 to the MV dye~\cite{zheng2007}. The valance band electrons in these nanoparticles under UV radiation with
  the photons of energy greater than or equal to ZnO band gap ($e^{-}$) can be excited to the conduction band
   producing the equal number of holes ($h^{+}$) in the valance band, simultaneously. Because the conduction band energy
   level of ZnO nanoparticles is higher than that of the Fermi level of ZnO:Ag nanoparticles, electrons can flow from ZnO 
   nanoparticles to Ag nanoparticles~\cite{kuriakose2014}.~So, oxygen vacancy defects and Ag nanoparticles on 
   the surface of ZnO nanoparticles trap electrons and prevents the recombination of $e^{-}-h^{+}$ pairs~\cite{chen2008}. 
   Also, UV radiation excites the dye molecules ($MV^{0}$ to $MV^{+}$).~These MV molecules transfer electrons
    to the conduction band (CB) of ZnO ($MV^{+}$ to CB of ZnO) and the Fermi level of Ag ($MV^{+}$ to Ag). 
   Then CB electrons react with dissolved oxygen in the solution to produce superoxide radicals($.O_{2}^{-}$),
    while the valance band (VB) holes react with hydroxide ions for the production of hydroxyl radicals($.OH^{-}$).
    Both of these radicals are responsible for the decolorization of MV dye~\cite{Kumar2015, kuriakose2014facile}.~
     Also, the energy barrier of the Ag-semiconductor junction prevented electron-hole recombination to separate photogenerated charges, effectively~\cite{Du2014}.

\subsection{Ab initio calculation of Ag-doped ZnO}
A 32 orthorhombic wurtzite supercell of ZnO:Ag 
(16 oxygen atoms, 15 Zinc atoms and an atom of silver substituting Zn)
corresponding to a nominal  Ag doping concentration of 6.25 $\%$ 
was studied by the density of states.
First principles calculations based on density functional theory were 
carried out using the HSE hybrid exchange-correlation functional at the 
level of HSE03~\cite{heyd2003}.
Sampling a 8$\times$8$\times$6 $\Gamma$-centered $k$-points mesh in the Brillouin zone 
was done for the calculation.
Hybrid functional correction in these calculations was seen to slightly reduce $E_{3d}$
from 6.98 eV in the density functional theory (DFT) ground state to 6.01 eV in the HSE, while forbidden gap was improved 
to 2.24 eV compared to 0.78 eV in the conventional GGA-DFT calculations.~

The present HSE calculations showed a deep level in the band gap of the single crystal  which 
explained the n-type conductivity due to Ag doping.
Hybridization between Ag-4d and O-2p levels led to this deep  
level above the valance band maximum (VBM).
Indeed, this deep level originated from Ag($d_{z^{2}}$)-O(2p) repulsion which pushed the 
Ag $d_{z^{2}}$ orbital to higher energies around 0.6 eV above VBM.
So, the existence of the deep levels was n-type conductivity
formation proof  in pure Ag-doped ZnO,
while grain boundaries in the powder samples can prepare shallow levels   
as acceptor p-type conductivity centers as found in these samples~\cite{wol2010}.
\begin{figure}
\begin{center}
\includegraphics*[scale=0.35]{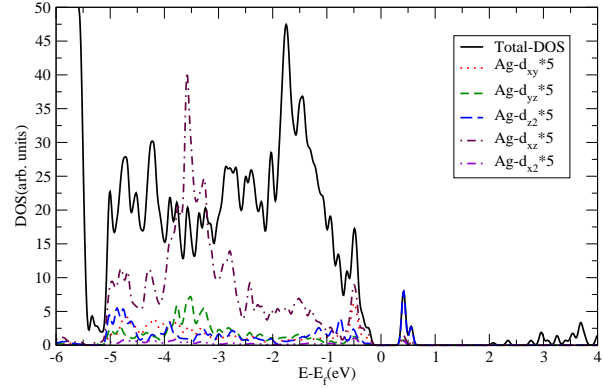}
\caption{\label{cal}
(Color online) The total and partial DOS of ZnO doped with Ag calculated via 
the HSE method.~Fermi level is set to zero.
}
\end{center}
\end{figure}

\section{Conclusion}
In conclusion, the optical, structural and photocatalytic properties of 
Ag-doped ZnO nanoparticles were checked.~XRD analysis showed that the samples were in the 
hexagonal-wurtzite phase.~Also, the XRD and XPS measurements confirmed the presence of silver 
metallic phase in the nanoparticles.~The density functional theory calculation predicted a deep level 
in the band gap of the single crystal which 
explained the n-type conductivity due to Ag doping.~In general, in the powder samples, the grain boundary 
and metallic Ag on the 
grain boundaries were found to prepare shallow levels as a smoking gun for p-type conductivity.
Optical absorption measurements indicated a red shift in 
the absorption band edge by silver doping.~This study suggested that silver doping reduced the
energy gap and controlled the luminescence properties; thus, it produced the p-type
conductivity.~This p-type nature can be used in the optoelectronic industry.~

The photocatalytic 
activity of the ZnO nanoparticles were improved by Ag doping that was confirmed by
XPS analysis.~Also, decolorisation
efficiency of methyl violet depended on the amount of silver.

\section{ACKNOWLEDGMENTS}
The authors would like to thank Dr.~Mehdi Ranjbar, Dr.~Mehran Ghiaci and Isfahan University of Technology for supporting this project.
\bibliographystyle{elsarticle-num}
\bibliography{ZnOAg}

\begin{thebibliography}{10}
\expandafter\ifx\csname url\endcsname\relax
  \def\url#1{\texttt{#1}}\fi
\expandafter\ifx\csname urlprefix\endcsname\relax\def\urlprefix{URL }\fi
\expandafter\ifx\csname href\endcsname\relax
  \def\href#1#2{#2} \def\path#1{#1}\fi

\bibitem{Das2008}
C.~D. Pemmaraju, R.~Hanafin, T.~Archer, H.~B. Braun, S.~Sanvito, Impurity-ion
  pair induced high-temperature ferromagnetism in co-doped zno, Phys. Rev. B 78
  (2008) 054428.
\newblock \href {http://dx.doi.org/10.1103/PhysRevB.78.054428}
  {\path{doi:10.1103/PhysRevB.78.054428}}.

\bibitem{Sarsari2013}
I.~A. Sarsari, C.~D. Pemmaraju, H.~Salamati, S.~Sanvito, Many-body
  quasiparticle spectrum of co-doped zno: A gw perspective, Phys. Rev. B 87
  (2013) 245118.
\newblock \href {http://dx.doi.org/10.1103/PhysRevB.87.245118}
  {\path{doi:10.1103/PhysRevB.87.245118}}.

\bibitem{tsu2004}
A.~Tsukazaki, A.~Ohtomo, T.~Onuma, M.~Ohtani, T.~Makino, M.~Sumiya, K.~Ohtani,
  S.~F. Chichibu, S.~Fuke, Y.~Segawa, et~al., Repeated temperature modulation
  epitaxy for p-type doping and light-emitting diode based on zno, Nature
  materials 4~(1) (2004) 42--46.

\bibitem{ozg2005}
{\"U}.~{\"O}zg{\"u}r, Y.~I. Alivov, C.~Liu, A.~Teke, M.~Reshchikov,
  S.~Do{\u{g}}an, V.~Avrutin, S.-J. Cho, H.~Morkoc, A comprehensive review of
  zno materials and devices, Journal of applied physics 98~(4) (2005) 041301.

\bibitem{ah2009}
M.~Ahmad, J.~Zhao, J.~Iqbal, W.~Miao, L.~Xie, R.~Mo, J.~Zhu, Conductivity
  enhancement by slight indium doping in zno nanowires for optoelectronic
  applications, Journal of Physics D: Applied Physics 42~(16) (2009) 165406.

\bibitem{tek2011}
S.~T. Teklemichael, W.~H. Oo, M.~McCluskey, E.~D. Walter, D.~W. Hoyt, Acceptors
  in zno nanocrystals, Applied Physics Letters 98~(23) (2011) 232112.

\bibitem{sahu2014}
D.~Sahu, N.~Panda, B.~Acharya, A.~Panda, Enhanced uv absorbance and
  photoluminescence properties of ultrasound assisted synthesized gold doped
  zno nanorods, Optical Materials 36~(8) (2014) 1402--1407.

\bibitem{yan2006}
Y.~Yan, M.~Al-Jassim, S.-H. Wei, Doping of zno by group-ib elements, Applied
  physics letters 89~(18) (2006) 181912--181912.

\bibitem{Vol2009}
Y.~Ma, G.~T. Du, S.~R. Yang, Z.~T. Li, B.~J. Zhao, X.~T. Yang, T.~P. Yang,
  Y.~T. Zhang, D.~L. Liu, Control of conductivity type in undoped zno thin
  films grown by metalorganic vapor phase epitaxy, Journal of Applied Physics
  95~(11).

\bibitem{thomas2012}
M.~Thomas, W.~Sun, J.~Cui, Mechanism of ag doping in zno nanowires by
  electrodeposition: experimental and theoretical insights, The Journal of
  Physical Chemistry C 116~(10) (2012) 6383--6391.

\bibitem{lupan2010}
O.~Lupan, L.~Chow, L.~K. Ono, B.~R. Cuenya, G.~Chai, H.~Khallaf, S.~Park,
  A.~Schulte, Synthesis and characterization of ag-or sb-doped zno nanorods by
  a facile hydrothermal route, The Journal of Physical Chemistry C 114~(29)
  (2010) 12401--12408.

\bibitem{Karun2011}
C.~Karunakaran, V.~Rajeswari, P.~Gomathisankar, Optical, electrical,
  photocatalytic, and bactericidal properties of microwave synthesized
  nanocrystalline ag-zno and zno, Solid State Sciences 13~(5) (2011) 923--928.

\bibitem{amo2012}
P.~Amornpitoksuk, S.~Suwanboon, S.~Sangkanu, A.~Sukhoom, N.~Muensit,
  J.~Baltrusaitis, Synthesis, characterization, photocatalytic and
  antibacterial activities of ag-doped zno powders modified with a diblock
  copolymer, Powder Technology 219 (2012) 158--164.

\bibitem{chen2008}
T.~Chen, Y.~Zheng, J.-M. Lin, G.~Chen, Study on the photocatalytic degradation
  of methyl orange in water using ag/zno as catalyst by liquid chromatography
  electrospray ionization ion-trap mass spectrometry, Journal of the American
  Society for Mass Spectrometry 19~(7) (2008) 997--1003.

\bibitem{Kumar2015}
R.~Kumar, D.~Rana, A.~Umar, P.~Sharma, S.~Chauhan, M.~S. Chauhan,
  \href{http://www.sciencedirect.com/science/article/pii/S0039914015000673}{Ag-doped
  zno nanoellipsoids: Potential scaffold for photocatalytic and sensing
  applications}, Talanta~(0) (2015) --.
\newline\urlprefix\url{http://www.sciencedirect.com/science/article/pii/S0039914015000673}

\bibitem{he2011}
M.~He, Y.~Tian, D.~Springer, I.~Putra, G.~Xing, E.~Chia, S.~Cheong, T.~Wu,
  Polaronic transport and magnetism in ag-doped zno, Applied Physics Letters
  99~(22) (2011) 222511.

\bibitem{li2012}
A.-Y. Li, X.-D. Li, Q.-B. Lin, S.-Q. Wu, Z.-Z. Zhu, Half-metallic
  ferromagnetism in ag-doped zno: An< i> ab initio</i> study, Solid State
  Sciences 14~(7) (2012) 769--772.

\bibitem{coey2005}
J.~Coey, M.~Venkatesan, C.~Fitzgerald, Donor impurity band exchange in dilute
  ferromagnetic oxides, Nature materials 4~(2) (2005) 173--179.

\bibitem{rad2013}
M.~S. Rad, A.~Kompany, A.~K. Zak, M.~Javidi, S.~Mortazavi, Microleakage and
  antibacterial properties of zno and zno: Ag nanopowders prepared via a
  sol--gel method for endodontic sealer application, Journal of nanoparticle
  research 15~(9) (2013) 1--8.

\bibitem{karunak2010}
C.~Karunakaran, V.~Rajeswari, P.~Gomathisankar, Antibacterial and
  photocatalytic activities of sonochemically prepared zno and ag-zno, Journal
  of Alloys and Compounds 508~(2) (2010) 587--591.

\bibitem{chau2010}
R.~Chauhan, A.~Kumar, R.~P. Chaudhary, Synthesis and characterization of silver
  doped zno nanoparticles, Archives of Applied Science Research 2~(5) (2010)
  378--385.

\bibitem{zandi2011}
S.~Zandi, P.~Kameli, H.~Salamati, H.~Ahmadvand, M.~Hakimi, Microstructure and
  optical properties of zno nanoparticles prepared by a simple method, Physica
  B: Condensed Matter 406~(17) (2011) 3215--3218.

\bibitem{Geo2008}
M.~K. P. S.~C. Georgekutty, Reenamole;~Seery, A highly efficient ag-zno
  photocatalyst: Synthesis, properties, and mechanism, The Journal of Physical
  Chemistry C 112.
\newblock \href {http://dx.doi.org/10.1021/jp802729a}
  {\path{doi:10.1021/jp802729a}}.

\bibitem{Kar2011}
C.~Karunakaran, V.~Rajeswari, P.~Gomathisankar, Combustion synthesis of zno and
  ag-doped zno and their bactericidal and photocatalytic activities,
  Superlattices and Microstructures 50~(3) (2011) 234--241.

\bibitem{Vo2009}
O.~Volnianska, P.~Boguslawski, J.~Kaczkowski, P.~Jakubas, A.~Jezierski,
  E.~Kaminska, Theory of doping properties of ag acceptors in zno, Phys. Rev. B
  80 (2009) 245212.
\newblock \href {http://dx.doi.org/10.1103/PhysRevB.80.245212}
  {\path{doi:10.1103/PhysRevB.80.245212}}.

\bibitem{Babu2014}
B.~Babu, T.~Aswani, G.~T. Rao, R.~J. Stella, B.~Jayaraja, R.~Ravikumar, Room
  temperature ferromagnetism and optical properties of cu< sup> 2+</sup> doped
  zno nanopowder by ultrasound assisted solid state reaction technique, Journal
  of Magnetism and Magnetic Materials 355 (2014) 76--80.

\bibitem{lanje2013}
A.~S. Lanje, S.~J. Sharma, R.~S. Ningthoujam, J.-S. Ahn, R.~B. Pode, Low
  temperature dielectric studies of zinc oxide (zno) nanoparticles prepared by
  precipitation method, Advanced Powder Technology 24~(1) (2013) 331--335.

\bibitem{Mur2014}
G.~Murtaza, R.~Ahmad, M.~Rashid, M.~Hassan, A.~Hussnain, M.~A. Khan,
  M.~Ehsan~ul Haq, M.~Shafique, S.~Riaz, Structural and magnetic studies on zr
  doped zno diluted magnetic semiconductor, Current Applied Physics 14~(2)
  (2014) 176--181.

\bibitem{Habibi2010}
M.~Habibi, R.~Sheibani,
  \href{http://dx.doi.org/10.1007/s10971-010-2177-x}{Preparation and
  characterization of nanocomposite zno–ag thin film containing nano-sized ag
  particles: influence of preheating, annealing temperature and silver content
  on characteristics}, Journal of Sol-Gel Science and Technology 54~(2) (2010)
  195--202.
\newblock \href {http://dx.doi.org/10.1007/s10971-010-2177-x}
  {\path{doi:10.1007/s10971-010-2177-x}}.
\newline\urlprefix\url{http://dx.doi.org/10.1007/s10971-010-2177-x}

\bibitem{AlG2013}
R.~Al-Gaashani, S.~Radiman, A.~Daud, N.~Tabet, Y.~Al-Douri, Xps and optical
  studies of different morphologies of zno nanostructures prepared by microwave
  methods, Ceramics International 39~(3) (2013) 2283--2292.

\bibitem{wahab2007}
R.~Wahab, S.~Ansari, Y.~Kim, H.~Seo, G.~Kim, G.~Khang, H.-S. Shin, Low
  temperature solution synthesis and characterization of zno nano-flowers,
  Materials Research Bulletin 42~(9) (2007) 1640--1648.

\bibitem{Jin2011}
Y.~Jin, Q.~Cui, K.~Wang, J.~Hao, Q.~Wang, J.~Zhang, Investigation of
  photoluminescence in undoped and ag-doped zno flowerlike nanocrystals,
  Journal of Applied Physics 109~(5) (2011) --.

\bibitem{Sah2012}
R.~K. Sahu, K.~Ganguly, T.~Mishra, M.~Mishra, R.~Ningthoujam, S.~Roy,
  L.~Pathak, Stabilization of intrinsic defects at high temperatures in zno
  nanoparticles by ag modification, Journal of colloid and interface science
  366~(1) (2012) 8--15.

\bibitem{Meng2013}
A.~Meng, S.~Sun, Z.~Li, J.~Han, Synthesis, characterization, and dispersion
  behavior of zno/ag nanocomposites, Advanced Powder Technology 24~(1) (2013)
  224--228.

\bibitem{Zheng2011}
J.~H. Zheng, J.~L. Song, X.~J. Li, Q.~Jiang, J.~S. Lian,
  \href{http://dx.doi.org/10.1002/crat.201100397}{Experimental and
  first-principle investigation of cu-doped zno ferromagnetic powders}, Crystal
  Research and Technology 46~(11) (2011) 1143--1148.
\newblock \href {http://dx.doi.org/10.1002/crat.201100397}
  {\path{doi:10.1002/crat.201100397}}.
\newline\urlprefix\url{http://dx.doi.org/10.1002/crat.201100397}

\bibitem{Chong2010}
M.~N. Chong, B.~Jin, C.~W. Chow, C.~Saint, Recent developments in
  photocatalytic water treatment technology: a review, Water research 44~(10)
  (2010) 2997--3027.

\bibitem{zheng2007}
Y.~Zheng, L.~Zheng, Y.~Zhan, X.~Lin, Q.~Zheng, K.~Wei, Ag/zno heterostructure
  nanocrystals: synthesis, characterization, and photocatalysis, Inorganic
  chemistry 46~(17) (2007) 6980--6986.

\bibitem{Ao2008}
Y.~Ao, J.~Xu, D.~Fu, C.~Yuan, Preparation of ag-doped mesoporous titania and
  its enhanced photocatalytic activity under uv light irradiation, Journal of
  Physics and Chemistry of Solids 69~(11) (2008) 2660--2664.

\bibitem{wu2010}
M.~Wu, B.~Yang, Y.~Lv, Z.~Fu, J.~Xu, T.~Guo, Y.~Zhao, Efficient one-pot
  synthesis of ag nanoparticles loaded on n-doped multiphase tio 2 hollow
  nanorod arrays with enhanced photocatalytic activity, Applied Surface Science
  256~(23) (2010) 7125--7130.

\bibitem{khosravi2014}
S.~Khosravi-Gandomani, R.~Yousefi, F.~Jamali-Sheini, N.~M. Huang, Optical and
  electrical properties of p-type ag-doped zno nanostructures, Ceramics
  International 40~(6) (2014) 7957--7963.

\bibitem{zheng2008}
Y.~Zheng, C.~Chen, Y.~Zhan, X.~Lin, Q.~Zheng, K.~Wei, J.~Zhu, Photocatalytic
  activity of ag/zno heterostructure nanocatalyst: correlation between
  structure and property, The Journal of Physical Chemistry C 112~(29) (2008)
  10773--10777.

\bibitem{thong2012}
K.~Thongsuriwong, P.~Amornpitoksuk, S.~Suwanboon, Photocatalytic and
  antibacterial activities of ag-doped zno thin films prepared by a sol--gel
  dip-coating method, Journal of sol-gel science and technology 62~(3) (2012)
  304--312.

\bibitem{pal2010}
B.~Pal, P.~K. Giri, High temperature ferromagnetism and optical properties of
  co doped zno nanoparticles, Journal of Applied Physics 108~(8) (2010) --.

\bibitem{mittal2014}
M.~Mittal, M.~Sharma, O.~Pandey, Uv--visible light induced photocatalytic
  studies of cu doped zno nanoparticles prepared by co-precipitation method,
  Solar Energy 110 (2014) 386--397.

\bibitem{jen2012}
T.-J. Whang, M.-T. Hsieh, H.-H. Chen, Visible-light photocatalytic degradation
  of methylene blue with laser-induced ag/zno nanoparticles, Applied Surface
  Science 258~(7) (2012) 2796--2801.

\bibitem{sinha2011}
M.~K. Gupta, N.~Sinha, B.~Kumar, p-type k-doped zno nanorods for optoelectronic
  applications, Journal of Applied Physics 109~(8) (2011) --.

\bibitem{Du2014}
L.~Zhang, L.~Du, X.~Yu, S.~Tan, X.~Cai, P.~Yang, Y.~Gu, W.~Mai, Significantly
  enhanced photocatalytic activities and charge separation mechanism of
  pd-decorated zno--graphene oxide nanocomposites, ACS applied materials \&
  interfaces 6~(5) (2014) 3623--3629.

\bibitem{liq2006}
J.~Liqiang, Q.~Yichun, W.~Baiqi, L.~Shudan, J.~Baojiang, Y.~Libin, F.~Wei,
  F.~Honggang, S.~Jiazhong, Review of photoluminescence performance of
  nano-sized semiconductor materials and its relationships with photocatalytic
  activity, Solar Energy Materials and Solar Cells 90~(12) (2006) 1773--1787.

\bibitem{zeferino2011}
R.~S. Zeferino, M.~B. Flores, U.~Pal, Photoluminescence and raman scattering in
  ag-doped zno nanoparticles, Journal of Applied Physics 109~(1) (2011) 014308.

\bibitem{yong2013}
Z.~Yong, C.~Xian, F.~Liguang, Y.~Lianfang, L.~Huihui, G.~Yuefei, Effects of
  annealing on the structural and photoluminescent properties of ag-doped zno
  nanowires prepared by ion implantation, Plasma Science and Technology 15~(8)
  (2013) 817.

\bibitem{Kar2012}
C.~Karunakaran, J.~Jayabharathi, K.~Jayamoorthy, P.~Vinayagamoorthy, Inhibition
  of fluorescence enhancement of benzimidazole derivative on doping zno with cu
  and ag, Journal of Photochemistry and Photobiology A: Chemistry 247 (2012)
  16--23.

\bibitem{Xu2013}
L.~Xu, G.~Zheng, J.~Miao, J.~Su, C.~Zhang, H.~Shen, L.~Zhao, Regulating effect
  of sio< sub> 2</sub> interlayer on optical properties of zno thin films,
  Journal of Luminescence 136 (2013) 307--312.

\bibitem{kohan2000}
A.~Kohan, G.~Ceder, D.~Morgan, C.~G. Van~de Walle, First-principles study of
  native point defects in zno, Physical Review B 61~(22) (2000) 15019.

\bibitem{Jay2013}
K.~S. R. S. K. C. V.~K. Jayakrishnan, R.;~Mohanachandran, Zno thin films with
  blue emission grown using chemical spray pyrolysis, Materials Science in
  Semiconductor Processing 16.
\newblock \href {http://dx.doi.org/10.1016/j.mssp.2012.10.003}
  {\path{doi:10.1016/j.mssp.2012.10.003}}.

\bibitem{Lin2013}
L.~Xu, G.~Zheng, H.~Wu, J.~Wang, F.~Gu, J.~Su, F.~Xian, Z.~Liu, Strong
  ultraviolet and violet emissions from zno/tio< sub> 2</sub> multilayer thin
  films, Optical Materials 35~(8) (2013) 1582--1586.

\bibitem{wang2004}
R.~Wang, J.~H. Xin, Y.~Yang, H.~Liu, L.~Xu, J.~Hu, The characteristics and
  photocatalytic activities of silver doped zno nanocrystallites, Applied
  Surface Science 227~(1) (2004) 312--317.

\bibitem{kuriakose2014}
S.~Kuriakose, V.~Choudhary, B.~Satpati, S.~Mohapatra, Enhanced photocatalytic
  activity of ag--zno hybrid plasmonic nanostructures prepared by a facile wet
  chemical method, Beilstein journal of nanotechnology 5~(1) (2014) 639--650.

\bibitem{kuriakose2014facile}
S.~Kuriakose, V.~Choudhary, B.~Satpati, S.~Mohapatra, Facile synthesis of
  ag--zno hybrid nanospindles for highly efficient photocatalytic degradation
  of methyl orange, Physical Chemistry Chemical Physics 16~(33) (2014)
  17560--17568.

\bibitem{heyd2003}
J.~Heyd, G.~E. Scuseria, M.~Ernzerhof, Hybrid functionals based on a screened
  coulomb potential, The Journal of Chemical Physics 118~(18) (2003)
  8207--8215.

\bibitem{wol2010}
W.~K\"orner, C.~Els\"asser, First-principles density functional study of dopant
  elements at grain boundaries in zno, Phys. Rev. B 81 (2010) 085324.
\newblock \href {http://dx.doi.org/10.1103/PhysRevB.81.085324}
  {\path{doi:10.1103/PhysRevB.81.085324}}.

\end{thebibliography}
\end{document}